\documentclass[iop]{emulateapj}
\usepackage{graphicx}
\usepackage{amsmath}
\usepackage{subfigure}
\usepackage{natbib}
\usepackage{txfonts}
\newcommand{\jms}{J.~Mol.~Spectr.}                %
\newcommand{\jpca}{J.~Phys.~Chem.~A}     %
\newcommand{\jpcrd}{J.~Phys.~Chem.~Ref.~Data}     %

\begin{document}
%
   \title{Unveiling the dust nucleation zone of IRC+10216 with ALMA}


\shorttitle{}
\shortauthors{J. Cernicharo et al.}

\author{J. Cernicharo\altaffilmark{1}, F. Daniel\altaffilmark{1}, A. Castro-Carrizo\altaffilmark{2}, M. Agundez\altaffilmark{3}, N. Marcelino\altaffilmark{4},
C. Joblin\altaffilmark{5,6}, J.R. Goicoechea\altaffilmark{1}, M. Gu\'elin\altaffilmark{2,7} }

\altaffiltext{1}{Departamento de Astrof\'isica, Centro de
Astrobiolog\'ia, CSIC-INTA, Ctra. de Torrej\'on a Ajalvir km 4,
28850 Madrid, Spain}
\altaffiltext{2}{Institut de Radioastronomie Millim\'etrique, 300 rue de la Piscine, 38406, Saint Martin d�H\`eres, France}
\altaffiltext{3}{Univ. Bordeaux, LAB, UMR 5804, 33270, Floirac, France}
\altaffiltext{4}{NRAO, 520 Edgemont Road, Charlottesville, VA22902, USA}
\altaffiltext{5}{Universit\'e de Toulouse; UPS-OMP; IRAP; Toulouse, France}
\altaffiltext{6}{CNRS, IRAP, 9 Av. col. Roche, 31028 Toulouse Cedex 4, France}
\altaffiltext{7}{LAM-LERMA, ENS, 24 rue Lhomond, 75005 Paris, France}

   \date{Received October, 10$^{th}$, 2013 ; accepted October, 25$^{th}$, 2013}

\begin{abstract}
We report the detection in IRC+10216 of lines of HNC $J$=3-2 
pertaining to 9 excited
vibrational states with energies up to $\sim$5300 K. The spectrum, observed with ALMA, also shows a surprising
large number of narrow, unidentified lines that arise in the vicinity of the star.
The HNC data are interpreted through a 1D--spherical non--local radiative transfer
model, coupled to a chemical model that
includes chemistry at thermochemical equilibrium for the innermost regions and reaction
kinetics for the external envelope.
Although unresolved by the present early ALMA data, the radius inferred for the emitting region 
is $\sim$0.06$\arcsec$ (i.e., $\simeq$ 3 stellar radii),
similar to the size of the dusty clumps reported by IR studies of the innermost region
($r <$ 0.3$\arcsec$).
The derived abundance of HNC relative to H$_2$ is $10^{-8} <$ $\chi$(HNC) $< 10^{-6}$, and
drops quickly where the gas density decreases and the gas chemistry is dominated by reaction kinetics.
Merging HNC data with that of molecular species present throughout the inner envelope, such
as vibrationally excited HCN, SiS, CS, or SiO, should allow us to characterize the physical and chemical
conditions in the dust formation zone.
\end{abstract}


\section{Introduction}
Whereas small molecular species form under a wide range of physical conditions,
the nucleation and growth of dust grains require high gas densities. In space, the 
main places for grain nucleation are the photospheres of AGBs stars, the ejecta
of SNe, and Wolf-Rayet stars \citep{Williams1987,Gail2010,Dwek2011}. The dust grains that form are expelled into the ISM, 
where they are processed by UV photons, energetic particles, and shocks.
Protoplanetary disks, which also harbor high density regions, have temperatures too 
low for efficient nucleation, but are favorable sites for coagulation processes that 
allow previouly formed dust grain seeds to grow from submicron diameters to millimetre and
centimetre size particles \citep{Blum2008} and, in proper conditions, to micro-planetesimals 
and planets.

Both astronomical observations and physical and chemical modeling are needed
to understand complex processes such as nucleation, dust processing,
and the interaction of dust with gas-phase molecules and atoms. The regions of grain nucleation 
are however restricted to a few stellar radii; heavily obscured, they are not observable at 
optical wavelengths. Early mid infrared observations  have
permitted to derive the physical properties of the circumstellar envelope (CSE) of IRC+10216 \citep{Keady1988,Keady1993,Ridgway1988,
Monnier2000}. However,
the closest AGB stars being more than one hundred parsecs away, mid and far infrared (IR), 
and submillimeter (sub-mm) observations have been limited so far by angular resolution. 
This is now changing with ALMA, which will be able to reach at sub-mm wavelengths resolutions comparable 
to stellar diameters. Although ALMA's present resolution and sensitivity are far from 
their nominal values, they outgrow those of any other sub-mm instrument. We present here the first 
sub-mm ALMA observations of an inner AGB star envelope, IRC+10216, 
and demonstrate their potential in revealing the dust formation zone.

Among the many molecular lines detected in the 20 GHz-wide spectrum, the lines of HNC, 
are fairly strong and can be traced from the immediate star surroundings to the outer CSE 
and have for our purpose a special interest. 
High--$J$ rotational lines of its ground vibrational level have recently 
been observed in IRC+10216 from {\it Herschel}/HIFI observations. These low resolution data (HPBW= 30$''$) 
allowed us to model the HNC abundance distribution down to a few tens of stellar 
radii from the star \citep{daniel2012}. The new HNC lines analyzed in this Letter not only 
were observed with a 50 times higher angular resolution, but pertain to excited vibrational states with
energies of several thousands kelvins, and probe a few stellar radii around the star.

\begin{figure*}
\begin{center}
\includegraphics[angle=0,scale=.67]{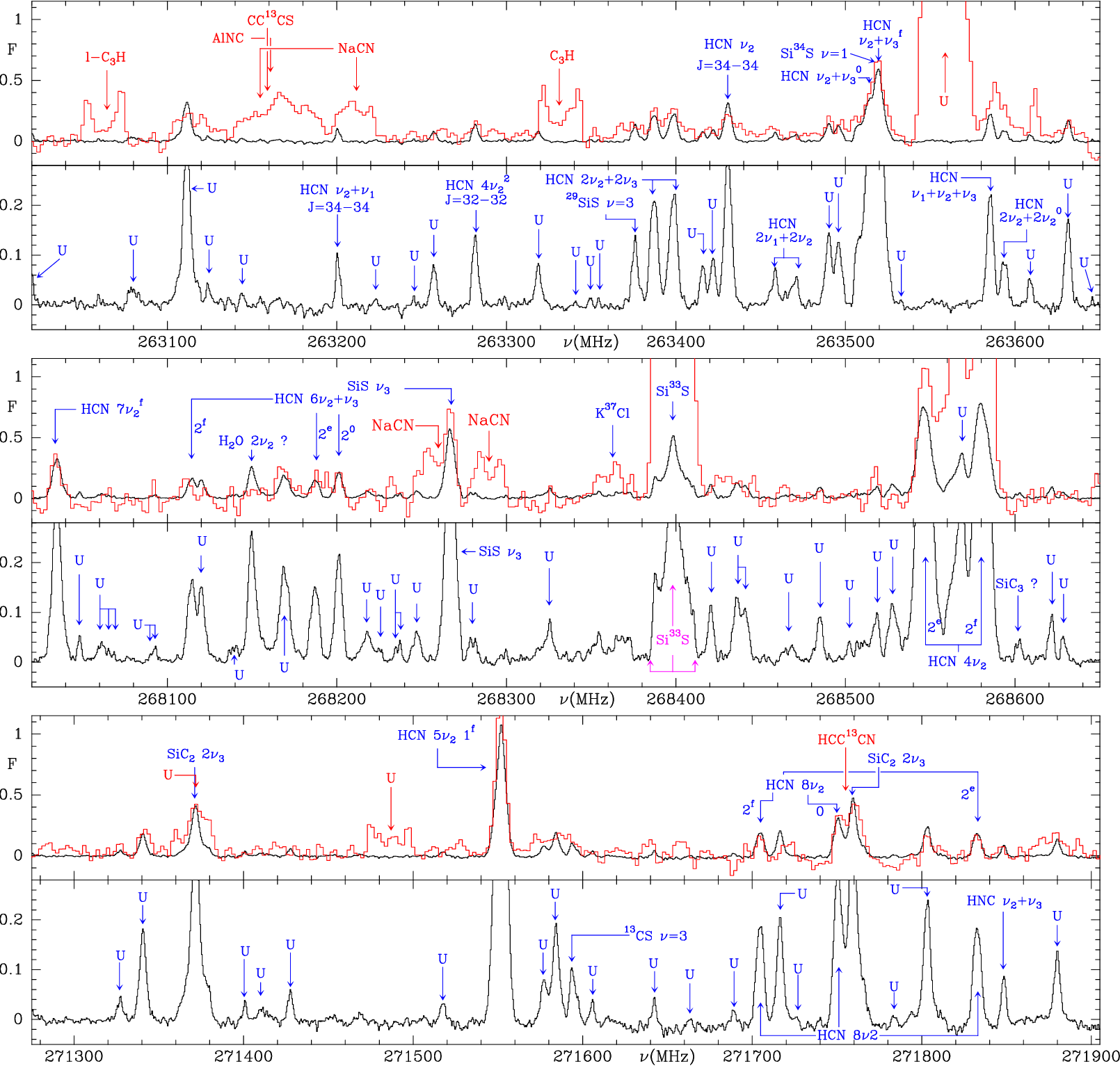} 
\caption{
ALMA spectrum (black) of IRC+10216 in three selected frequency ranges of the 20 GHz bandwidth 
covered by our data. For each selected frequency the ALMA spectrum is compared to
the data obtained with the 30-m IRAM telescope (red) at the same frequency
\citep{Cernicharo2011}. Intensity scale is in units of Jy/Beam. Spectral resolution is $\sim$1\,MHz for
both datasets.
The bottom panel 
of each selected frequency shows a close-up view of the ALMA data.
Interestingly, a forest of narrow and unidentified lines are
evidenced thanks to the extreme sensitivity, and much higher angular
resolution, of ALMA. Note the good
calibration agreement for lines spatially unresolved by both instruments. Labels in red
correspond to lines detected with the 30-m telescope and filtered by the interferometer.} \label{fig1}
\vspace{-0.5cm}
\end{center}
\end{figure*}

\begin{figure*}
\begin{center}
\includegraphics[angle=0,scale=.60]{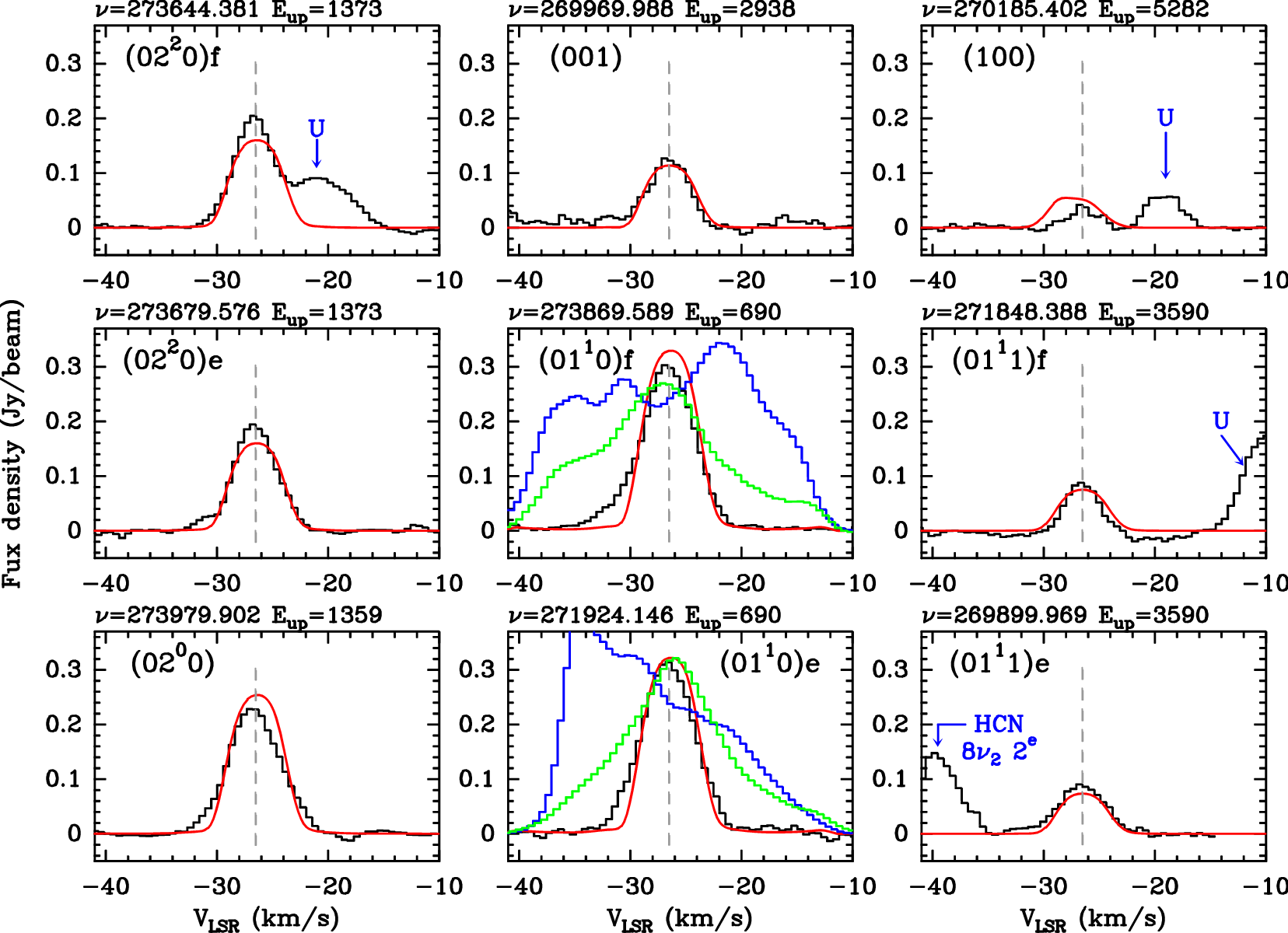} 
\caption{Observed (black curves) and modeled (red) line profile of the $J=$3-2
in vibrationally excited states of HNC as observed with ALMA near $\sim$270 GHz.
The blue and green curves show the
corresponding $\nu_2$ lines of HCN (blue) and H$^{13}$CN (green) observed in the same setup, scaled down by factors of 50 and 6 respectively.
HNC frequencies (in MHz) and upper level energies (in kelvins) are given at the top of each box. The vertical line
corresponds to the systemic velocity of IRC+10216.} \label{fig2}
\vspace{-0.5cm}
\end{center}
\end{figure*}

\section{Observations}

The observations of IRC+10216 with ALMA were made between April 8 and 23, 2012
(ALMA Cycle0). They consisted in 4 spectral setups covering the frequency
range 255.3 to 274.8\,GHz (receiver band 6) with a 0.5\,MHz channel spacing. 
\footnote{This paper makes use of the following ALMA data:
ADS/JAO.ALMA2011.0.00229.S. ALMA is a partnership of ESO (representing its
member states), NSF (USA) and NINS (Japan), together with NRC (Canada) and
NSC and ASIAA (Taiwan), in cooperation with the Republic of Chile. The Joint
ALMA Observatory is operated by ESO, AUI/NRAO and NAOJ}.
Observations were obtained in 2 runs in a single spectral setup
(referenced as id4 in the ALMA archive), which covered the range 269.8
to 274.0GHz. The array was configured with 16 antennas extending up to
402 m.  Two runs of about 72 min each were performed, of which 26 min
correspond to correlations on source. 3C279 and 3C273 were observed to
calibrate bandpass and absolute flux, J0854+201 and J0909+013 every 10
and 20 min respectively to calibrate phase and amplitude gains.  3C279
was bright enough to allow an optimal radio frequency (RF)
calibration.  A flux of 4.26 Jy was adopted as reference for J0854+201
at 272.4 GHz (consistent with SMA and PdBI monitoring of this
source). We estimate an uncertainty for the flux calibration of $\sim$
8\%, which is obtained from the reference flux uncertainty and the
resulting differences in source continuum for the different observing
runs. In addition, this standard calibration was subsequently improved
by iterative self-calibration on the highly compact, bright source at
the origin of vibrationally excited SiS emission (SiS $\nu$=1 J=15-14 at
270.917 GHz).  Imaging restoration was made with moderate robust
weighting (with a 0.5 robustness value) for all the available
channels, with 0.49 MHz spacing and an effective resolution of
0.98MHz.  The final synthetic beam is 0.6"x0.5" and the r.m.s. per
channel is $\sim$5\,mJy/Beam. That of the continuum map is 2\,mJy/Beam
with the continuum emission peak at $\alpha$=09$^h$\,47$^m$\,57.446$^s$ and
$\delta$=13$^o$\,16'\,43.86" (J2000). Calibration was performed
with the CASA software package, and data analysis with GILDAS.  We
finally note that in the originally delivered raw data (in ASDM
format) the frequency plane was wrong by 2.89 MHz for the spectral
window 1.  When the error was identified, it could be simply corrected
in the ms-format data files. We expect this error to be corrected in
the ALMA archive data.
%
%
%
%

\section{Line modeling and Discussion}

Figure \ref{fig1} shows for three selected frequency ranges the ALMA spectrum superimposed on the 
spectrum observed with the IRAM 30-m single-dish telescope \citep{Cernicharo2011}.
The line identification was made with the MADEX code \citep{Cernicharo2012}. 
The spectra show rotational lines of HCN, HNC, SiS (up to $\nu$=10), SiO, CS, SiC$_2$,
and metal-bearing species, together with their rare isotopologues, in the ground and excited vibrational states. 
Many pure $\ell$-doubling lines ($\Delta$J=0) of HCN in several vibrational levels are detected (see Figure \ref{fig1};
see also \citet{Patel2011}).
In addition, the ALMA spectrum shows an impressively large number of unidentified lines. The line width of the observed 
features ranges from 2-3\,km\,s$^{-1}$ to 10-15\,km\,s$^{-1}$. Narrow lines are formed close to 
the star's photosphere before the acceleration region. 
The line width of the features increases when the emission arises from
the acceleration region \citep{Fonfria2008, Cernicharo2011, Patel2011}.
A significant number of unidentified lines agree in frequency with lines of vibrationally excited CH$_2$CHCN, a molecule
known to be present in the outer envelope of IRC+10216 \citep{Agundez2008}.
However, a conclusive assignment awaits for a wider frequency coverage to access all major CH$_2$CHCN lines,
as well as for a more detailed model of the innermost CSE chemistry \citep{Agundez2006,Agundez2010}. A narrow feature at 268148 MHz agrees
in frequency with the 2$\nu_2$ 6$_{5,2}$-7$_{4,3}$ line of water vapor. A final assignment
of this high energy transition requires the detection of more features of vibrationally excited
H$_2$O; so far, only lines from the ground H$_2$O vibrational state have been reported 
in IRC+10216 \citep{Decin2010}. 

In this work we focus on the emission of the HNC lines shown in Figure \ref{fig2} and arising
from different vibrational modes.  
Their frequencies and intrinsic strengths have been calculated with
the MADEX code \citep{Cernicharo2012} from a fit of all laboratory data, including the
$\ell$--type doubling lines from $\nu_2$ levels
\citep{Okabayashi1993,Creswell1976,Thorwirth2000,maki2001,Mellau2010,Mellau2011,Amano2002,Amano2004}.

\subsection{Radiative Transfer and Chemical Modeling} 
The different gas phase processes leading to the formation of molecules throughout the CSE of IRC+10216 
were reported by \citet{Agundez2006}. The chemistry 
of HNC in the outer CSE of IRC +10216 was extensively discussed by \citet{daniel2012}. We therefore focus here on the chemistry in the 
very innermost region, i.e. within a few stellar radii from the star where thermochemical equilibrium ({\it TE}) holds. 
We carried out {\it TE} calculations with the code described
in \citet{tejero1991}, which is based on the formalism of \citet{tsuji1973}. We adopted solar elemental abundances
\citep{asplund2009}, except for C, whose abundance was increased over O by a factor of 1.5.
Our calculations include 21 elements and nearly 300 molecular species, whose thermochemical properties
were mostly taken from various compilations \citep{chase1998,sauval1984,irwin1988}. For HNC, the dissociation energy
and partition function were taken from the more recent studies of \citet{wenthold2000} and \citet{barber2002},
respectively. The {\it TE} equations were solved for the region 1-10 R$_*$ of IRC+10216,
adopting the temperature radial profile of  \citet{agundez2012}. A radial profile of H$_2$ volume density somewhat 
lower than that of \citet{agundez2012} was adopted to adequately reproduce the HNC observed line profiles (see below).

\begin{figure}
\begin{center}
\includegraphics[angle=0,scale=.60]{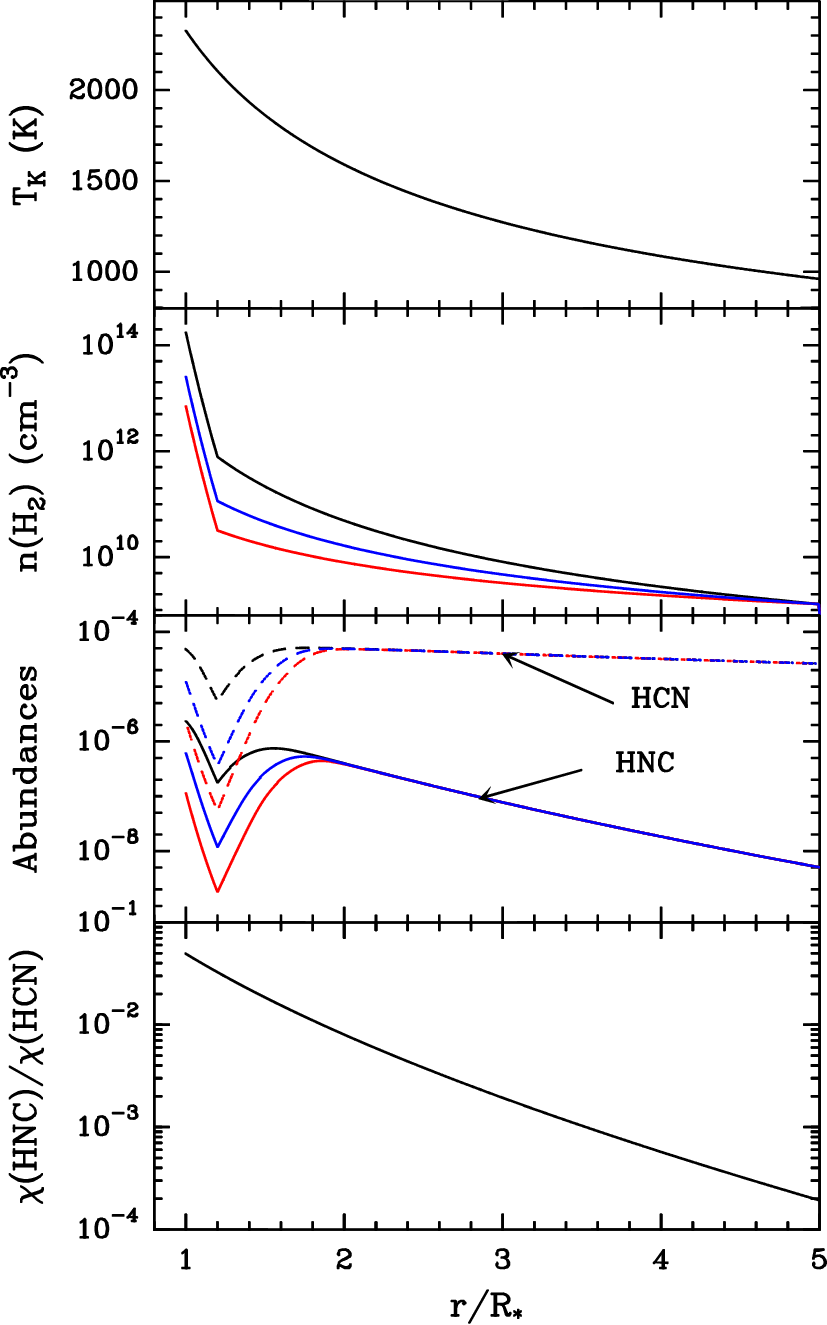}
\caption{
From top to bottom : gas temperature, H$_2$ density,
HCN \& HNC fractional abundance, and the HNC/HCN abundance ratio
adopted in our models.
With the gas temperature fixed in
the chemical modeling, the effect introduced by varying the H$_2$ density 
on the HNC/HCN fractional abundances is emphasized by considering three different density radial profiles.
The best results are obtained for the blue profile. The HNC/HCN abundance ratio does not depend on
the adopted density profile.} \label{fig3}
\vspace{-0.5cm}
\end{center}
\end{figure}

\begin{figure}
\begin{center}
\includegraphics[angle=-90,scale=.36]{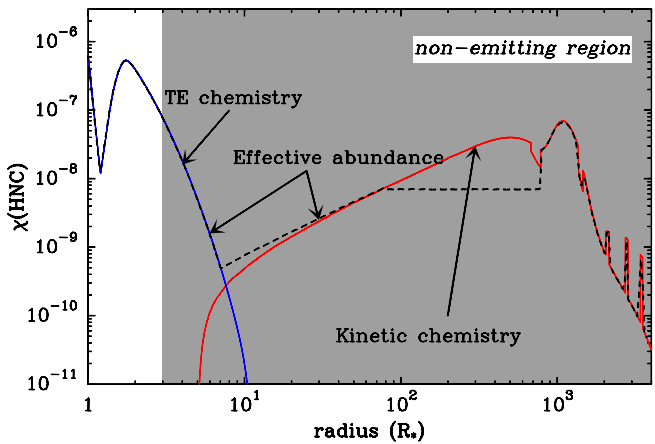}
\caption{Radial profile of the HNC abundance, relative to H$_2$, as predicted from the {\it TE} chemistry (blue curve)
and from the kinetic chemistry (red curve).
The effective abundance profile derived by \citep{daniel2012} from at fit to all the sub-mm lines observed by {\it Herschel} 
is shown by the dashed black curve. The gray area corresponds
to the region that does not contribute to vibrationally excited HNC emission.} \label{fig4}
\vspace{-0.5cm}
\end{center}
\end{figure}

Since HCN is one of the 
most abundant molecules in the atmosphere of carbon-rich AGB stars such as IRC+10216 
\citep{Fonfria2008,Cernicharo1996,Cernicharo1999,Cernicharo2011}, 
HNC must also be fairly abundant near the photosphere.
The calculated abundance of HNC is shown on Figures \ref{fig3} \& \ref{fig4} (blue line). It is maximum at the stellar surface 
(2 $\times$ 10$^{-6}$ relative to H$_2$) and drops steeply with increasing radius. At {\it TE},
the HNC/HCN abundance ratio increases exponentially with temperature and depends only on that parameter. 
It reaches a value of $\sim$10 \% for a photospheric temperature of 2500\,K \citep{Cernicharo2011}. 

According to the above considerations, the fractional HNC abundance, $\chi$(HNC), can be estimated in the {\it TE} 
regime if the gas temperature, HCN abundance, and volume density are known. \citet{Fonfria2008} and \citet{Cernicharo2011} have 
shown that T$_K$ and $\chi$(HCN) are well constrained by the HCN observations in the mid-IR and sub-mm domains.
This is not the case, however for the H$_2$ volume density (see \citealt{agundez2012}). 
On the assumption that in the innermost regions the temperature radial profile is well known and {\it TE} holds, the HNC line 
intensities observed by ALMA can be used to constrain the H$_2$ volume density in these regions.

In order to constrain the abundance of HNC, we performed
non--local radiative transfer (RT) calculations with the code described in \citet{daniel2008}.
Test calculations showed that the predicted line intensities depend
on the total number of vibrational levels included in the calculations. The dipole moments of the
vibrational bands were taken from
\citet{maki2001}. If not available, they were assumed identical to those of
transitions with similar variations of the vibrational quantum numbers. For instance, the dipole moments of the
$(03^10) \to (01^10)$ and $(02^00) \to (000)$ bands, were assumed to be the same.
The vibrational states which are included in the RT calculations are the
(000), (010), (020), (030), (100), (001), (011), (021). In each vibrational state
we consider the rotational levels up to $J$=55 and include the $\ell$--type doubling in the states
with a bending mode. The number of energy levels considered is thus $\sim$950, which translates to
a total of $\sim$8700 radiative transitions.

The collisional rate coefficients for the rotational transitions of HNC
were derived from the calculations of \citet{dumouchel2010}. Those 
for the ro-vibrational and $\ell$-type transitions are not listed in the literature. They were estimated by scaling down the 
rotational rate coefficients by a factor of 100, a value similar to that found for SiO at T$_K$=1500 K
\citep{Bieniek1983}. Several runs were performed to check the effect of this assumption
on the results. We found that for the range of considered densities, the pumping of the molecular levels
was not very sensitive to the assumed factor. Moreover, we tested that the pumping by IR photons is not important for these
densities and for the considered vibrational levels.

We adopt a distance to IRC+10216 of 130 pc \citep{groenewegen2012} and a radius for the
star of 4$\times$10$^{13}$\,cm \citep{menten2012}.
The structure of the CSE, i.e. its gas/dust temperatures, dust opacity,
and radial velocity profiles are taken from \citet{Fonfria2008}, \citet{agundez2012} and \citet{daniel2012} (see also
\citet{Keady1988,Ridgway1988,Keady1993,Agundez2006,Agundez2010,Cernicharo2010,debeck2012}).
While the HCN/HNC abundance ratio is expected to depend on temperature, the 
chemical modeling shows that $\chi$(HNC) is largely influenced by the adopted H$_2$ density for r$\leq$2\,R$_*$ (c.f. Figure \ref{fig3}).
Nevertheless, the observed HNC intensities are directly proportional to the column density of HNC.
Hence, by using the HCN abundance profile derived by \citet{Fonfria2008}, and adopting the {\it TE} HCN/HNC abundance
ratio (c.f. Figure \ref{fig3}), it is possible to constrain n(H$_2$) from the observations and the models.
The best fit to the data is obtained with the density profile shown by
the blue curves in Figure \ref{fig3}. It corresponds to a downward revision of the density profile of \citet{agundez2012}, 
and reproduces reasonably well the observations (see Figure \ref{fig2}).

\citet{daniel2012} have already analyzed several HNC lines observed by {\it Herschel} and tried to constrain
abundance profile of HNC over the entire envelope (see Figure \ref{fig4}).
They found that gas-phase chemistry may reproduce this profile, except for the $70 \, R_* < r < 700 \, R_*$ region, 
where the gas-phase predicted abundances are significantly lower than observed (see Figure \ref{fig2}), a discrepancy that
was explained by the large uncertainty on the HCNH$^+$ + NH$_3$ branching ratio.

In the current modeling, we kept for radii $\geq 10 \, R_*$ the $"$effective$"$ abundance profile derived from the analysis 
of the {\it Herschel} observations by \citet{daniel2012}, but adopted for smaller radii the abundances predicted by {\it TE} 
since the signal from this region is highly diluted in the {\it Herschel} beam and can not be modeled properly. 
Thus, the effective HNC abundance profile used in this study to fit all the HNC lines ({\it Herschel}, 30-m telescope and ALMA) follows 
the black dashed curve on Figure \ref{fig4} down to $\sim 10 \, R_*$ and the blue curve closer to the star.
With such a profile, the RT calculations show that the vibrationally excited HNC lines observed by ALMA 
are only sensitive to the hot $1 \, R_* < r < 3 \, R_*$ region, where the HNC abundance is expected to be at TE. And indeed, 
Figure \ref{fig2} illustrates that the {\it TE} abundances correctly predict the observed line profiles -- except for a
weak blue wing visible on the two $\ell$--type doubling components of the lowest excited vibrational state (0,1,0). The latter 
must correspond to expanding gas in front of the star, as it disappears in all the higher energy lines. Figure \ref{fig2} also 
shows the two J=3-2 $\nu_2$ $\ell$-doubling lines of HCN and H$^{13}$CN
for comparison. Those lines, especially those of the abundant isotopologue, are much broader and extend over the full velocity 
range of the envelope, i.e. trace the entire gas acceleration region, $1 \, R_* < r < 20 \, R_*$ \citep{Fonfria2008,Cernicharo2011}. 
In contrast, vibrationally excited 
HNC lines only trace the innermost region ($r < 3 \, R_*$).

\subsection{Discussion}

The structure of the innermost region of
IRC+10216 has been studied through interferometric observations of the dust near-infrared emission
\citep[see e.g.][]{menshchikov2001,leao2006}, but is still a matter of debate.
While the CSE seems to have a roughly spherical shape at
large scales, the region 
$r < 0.3 \arcsec$ ($R_*$=0.02 arcseconds)  
is dominated by a few bright clumps \citep{haniff1998}
with diameters of typically $\sim$0.1$\arcsec$ , whose
position and luminosity change on time--scales of a few years \citep[see e.g. Figure 2 of][]{leao2006}. 
Although these clumps are not resolved in our ALMA data, their size  of $\sim$0.12$\arcsec$
in diameter is close to that of the HNC emitting region.
The position of the star is not firmly established at this scale. Indeed, its location is chosen either
to coincide with the brightest clump \citep{tuthill2005}, or to lie in the middle of the clumps.
It may be also possible that the star is hidden at the wavelengths at which the clumps were observed.
\citep{menshchikov2001,menshchikov2002}. 
The clumps have projected velocities ranging from $\sim$8 km/s to 18 km/s
\citep{tuthill2000}. Hence, it should be possible,
in principle, to locate the star position by constraining the kinematics of the various clumps.
However, in practice, this turns out to be a difficult task because of the proper motion of the star 
\citep{menten2012}. The possibility  that IRC+10216 hosts in fact a binary star \citep{Guelin1993} has also 
to be addressed with more sensitive dust and molecular emission observations.

With our current model, it has been possible to successfully reproduce the emission of all the vibrationally excited 
$J$=3-2 HNC lines up to energy levels $\simeq  5300\,$K. The derived structure (assumed spherically symmetrical) implies 
a hot and dense emitting region ( $n$(H$_2$) $>$ $5\, 10^9$ cm$^{-3}$ and T$_g$ $>$ 1300 K, in our model) extending 
from 1\,$R_*$ up to $3\,R_*$ (corresponding to a projected distance of $\sim$0.06$\arcsec$ at 130 pc). 
In a more realistic non spherically 
symmetric model, and taking into account episodic events of high mass loss, 
those lines would arise from the dense clumps and not from a less dense 
homogeneous interclump medium.
We have shown that for densities much lower than $10^9$ cm$^{-3}$, the HNC chemistry is dominated by bimolecular 
chemical reactions making its fractional abundance to drop below $\chi$(HNC)$ < 10^{-8}$, a value too 
low to explain the observed line strengths. 

The strong variation of the abundance of HNC with distance to the star is illustrated on Figure \ref{fig2} by the
comparison between the HNC and the HCN and H$^{13}$CN line profiles. The latter lines
are 50 and 6 times stronger, respectively, at our angular resolution, and are seen to cover 
the entire gas acceleration region \citep{Cernicharo2011}.
The emitting region is still unresolved with the current observations ($\simeq$0.6$\arcsec$ synthetized beam) 
and an angular resolution better than $\sim$0.05$\arcsec$, as well as 2D or 3D modeling would be required to 
fully characterize this region. 

HNC and HCN are excellent molecular tracers as their chemistry is strongly correlated in AGB stars.
Coupling chemistry and radiative transfer modelling will be a powerful tool to study the innermost
regions of CSEs. Higher angular resolution observations of HNC and HCN, that will become possible with ALMA in the 
coming years, and their analysis in conjuction with other dense refractory gas tracers, such as SiO, CS, SiC$_2$, SiS, 
and metal-bearing species, will permit to further study
the morphology and physics of the dust formation/gas acceleration zone and, mostly, allow us to follow 
on time-scale of years the time-dependant chemistry in the moving gas clumps. 
This will open the possibility to study the elemental processes that lead to the nucleation and growth of refractory dust grains

\acknowledgements
We would like to thank Spanish MINECO for funding support under grants CSD2009-
00038, AYA2006-14876, AYA2009-07304, AYA2012-32032, the CNRS/PCMI program,
and the RTRA STAE (3PC project).



\end{document}